\documentclass[12pt]{article}

\usepackage{amsmath,amssymb,amsfonts}
\usepackage{graphicx}
\usepackage{booktabs}
\usepackage[margin=1in]{geometry}
\usepackage{hyperref}
\graphicspath{{figures}}
\usepackage{float}

\title{First-return statistics in bounded radiative transport: A Motzkin polynomial framework}

\author{Claude Zeller\\
Claude Zeller Consulting LLC, Tillamook, Oregon 97134\\
\texttt{czeller@ieee.org}
\and
Robert Cordery\\
Fairfield University, 1073 North Benson Rd. Fairfield CT 06824\\
\texttt{rcordery@fairfield.edu}}

\date{}

\begin{document}

\maketitle

\begin{abstract}
A photon entering a scattering medium executes a three-dimensional random walk determined by the Henyey--Greenstein phase function. The photon either reaches the boundary for a first passage or is absorbed. Projecting the walk onto the axial direction produces a one-dimensional alternating process whose peaks and valleys correspond to changes in the sign of the projected step. This reduction preserves first-return and first-passage events and leads to a representation in terms of Motzkin-type polynomials.

The analytical formulation is complete except for boundary-constrained return terms, which appear as high-order integrals. We treat these contributions with a single truncation factor determined from Monte Carlo simulations of first-return distributions over a wide range of anisotropy \(g\) and scattering steps \(m_s\). The resulting factor follows a Cauchy distribution. Incorporating it yields first-return probabilities in agreement with full three-dimensional Monte Carlo to within 2\% for \(g \le 0.7\).

The approach gives backscattering coefficients from phase-function integrals and provides an efficient alternative to full three-dimensional simulations for problems of radiative transport in semi-infinite media.
\end{abstract}


\vspace{1em}
\noindent\textbf{Principal notation:}
\begin{center}
\small
\begin{tabular}{ll@{\qquad}ll}
$g$ & anisotropy factor & $m_s$ & number of scattering steps \\
$\mu$ & cosine of scattering angle & $\mu_b$ & forward/backward threshold \\
$r_b$ & effective backscattering coefficient & BTF & Boundary Truncation Factor \\
$\gamma$ & dimensional reduction exponent & $M_n(t)$ & Motzkin polynomial \\
$T(n,k)$ & Motzkin triangle coefficient & $P_\mathrm{HG}$ & Henyey--Greenstein phase function \\
\end{tabular}
\end{center}

\section{Introduction and background}

We address the canonical radiative transport problem: normally incident illumination on a plane boundary of a homogeneous semi-infinite medium with exponential free-path statistics and Henyey--Greenstein phase function scattering~\cite{HenyeyGreenstein1941}. This configuration serves as a fundamental benchmark for testing asymptotic approximations and numerical methods in radiative transfer theory~\cite{Chandrasekhar1960}.

The problem's apparent simplicity belies its analytical difficulty. While the Henyey--Greenstein phase function admits elegant compositional properties under convolution, boundary constraints on first-return statistics introduce high-dimensional integrals that resist closed-form evaluation beyond the first few scattering orders. Existing approaches---diffusion approximations, Kubelka--Munk two-flux models~\cite{KubelkaMunk1931}, discrete ordinate methods---each sacrifice mathematical rigor for tractability, leaving the field without a controlled analytical-numerical benchmark.

Existing extensions of Kubelka--Munk to three dimensions have not resolved this difficulty. A decade ago, Sandoval and Kim~\cite{SandovalKim2014} derived a generalized Kubelka--Munk (gKM) approximation using double spherical harmonics, representing the most systematic such extension to date. However, their method fails for forward-peaked scattering---producing unphysical negative intensities at anisotropy factors typical of biological tissue---a limitation that has persisted and excludes most applications in tissue optics and remote sensing.

This paper develops such a benchmark through three contributions. First, we introduce a dimensional reduction mapping the 3-D anisotropic random walk onto a 1-D Motzkin polynomial framework~\cite{OsteVanderJeugt2015}, recognizing that the axial projection constitutes a hidden Markov process. Second, we define the Boundary Truncation Factor (BTF) to handle the analytically intractable integrals that arise at higher scattering orders. Third, we calibrate and validate the framework against extensive Monte Carlo simulations~\cite{Jacques2010,Sassaroli1998}, revealing unexpected regularity in BTF behavior.

\subsection{Applications}

The framework addresses transport calculations in three domains. In biomedical optics, tissue reflectance spectroscopy requires accurate first-return statistics for optical property extraction~\cite{JacquesTissue2013}. In atmospheric physics, radiative transfer through cloud layers involves strongly forward-peaked scattering where diffusion approximations fail~\cite{Chandrasekhar1960}. In materials science, appearance modeling demands transport solutions that preserve boundary effects efficiently~\cite{Modric2009}. First-return problems also arise in financial applications where path-dependent statistics govern option pricing and risk assessment~\cite{Ballestra2016}.

\subsection{Paper organization}

Section~2 reviews one-dimensional first-passage theory and establishes the Motzkin polynomial formalism. Section~3 presents the Henyey--Greenstein phase function properties essential for dimensional reduction. Section~4 develops the Boundary Truncation Factor framework and presents the complete computational algorithm. Section~5 describes Monte Carlo validation. Section~6 presents conclusions and discusses limitations.

\section{One-dimensional first-passage theory}

The Kubelka--Munk (KM) theory~\cite{KubelkaMunk1931}, initially developed in 1931, has undergone significant evolution~\cite{Haney2007,Myrick2011}. Yang and Kruse~\cite{YangKruse2004} introduced path-length corrections due to scattering effects. Sandoval and Kim~\cite{SandovalKim2014,SandovalKim2017} derived KM from the rigorous radiative transfer equation and extended it to three dimensions and polarization. Philips-Invernizzi et al.~\cite{PhilipsInvernizzi2001} provide a comprehensive review of earlier work on diffusing media reflectance. Several researchers have explored improved statistical methods for first-passage problems~\cite{CondaminBenichou2005,BlancoFournier2006,Libois2022}. Asymptotic properties have been of particular interest~\cite{BorovkovBorovkov2008,McDonald1999}. Two comprehensive monographs dealing with random walks and first-passage problems are Redner~\cite{Redner2001} and Rudnick and Gaspari~\cite{RudnickGaspari2004}.

\subsection{Radiative transfer and the KM equation}

The Kubelka--Munk model solves the one-dimensional radiative transfer problem~\cite{Chandrasekhar1960,Chandrasekhar1943}:
\begin{align}
\frac{dI_f}{dz} &= -\chi I_f(z) + S I_b(z) \\
\frac{dI_b}{dz} &= \chi I_b(z) - S I_f(z)
\end{align}

With scattering rate $S$ and absorption coefficient $\chi$, the KM reflectance from a semi-infinite medium is:
\begin{equation}
R_\infty(S,\chi) = \frac{S+\chi}{S} - \sqrt{\left(\frac{S+\chi}{S}\right)^2 - 1}
\label{eq:km_reflectance}
\end{equation}

The reflectance is the Laplace transform of the first-passage path-length distribution:
\begin{align}
R_\infty(S,\chi) &= \mathcal{L}_\lambda\{P(\lambda;S)\}(\chi) = \int_0^\infty P(\lambda;S) e^{-\chi\lambda}\, d\lambda \\
P(\lambda;S) &= \mathcal{L}_\chi^{-1}\{R_\infty(S,\chi)\}(\lambda)
\end{align}

\subsection{Catalan numbers and combinatorial structure}

The inverse Laplace transform reveals that $R_\infty$ is proportional to $C(x)$, the generating function for the Catalan numbers~\cite{Sloane2023}:
\begin{equation}
C(x) = \sum_{n=0}^{\infty} C_n x^n = \frac{1 - \sqrt{1-4x}}{2x}
\end{equation}
\begin{equation}
R_\infty(S,\chi) = \frac{1}{2} \cdot \frac{S}{S+\chi} \cdot C\left(\frac{S^2}{4(S+\chi)^2}\right)
\end{equation}
where $C_n = \frac{(2n)!}{n!(n+1)!}$. Previous work established that Catalan numbers, which describe the combinatorics of Dyck paths on a lattice, apply to random walks in scattering media~\cite{DrakeGantner2011,SimonTrachsler2003}.

\subsection{Motzkin polynomials: extension to forward scattering}

\textbf{Limitation of Catalan numbers.} Catalan numbers describe Dyck paths---random walks restricted to up- or down-steps corresponding to complete directional reversals at each scattering event. This constraint applies to pure backscattering but fails to represent forward scattering bias where photons maintain preferential forward propagation.

\textbf{Motzkin extension.} Forward scattering requires flat steps in combinatorial structures---events where photons continue forward without $z$-coordinate projection changes. This necessitates Motzkin paths~\cite{OsteVanderJeugt2015,DrakeGantner2011}, which extend Dyck paths through horizontal moves alongside up and down steps.

The one-dimensional first-return probability extends to~\cite{ZellerCordery2020}:
\begin{equation}
P_\mathrm{refl}^{(1D)}(m_s, r) = (1-r)^{m_s-1} \sum_{n_p=1}^{\lfloor m_s/2 \rfloor} T(m_s-2, n_p-1) \left[\frac{r}{2(1-r)}\right]^{2n_p-1}
\label{eq:1d_reflection}
\end{equation}

Compact representation using Motzkin polynomial $M_{m_s-2}(t)$:
\begin{equation}
P_\mathrm{refl}^{(1D)}(m_s, r) = M_{m_s-2}\left(\frac{1-2r}{r}\right) \cdot \left(\frac{r}{2}\right)^{m_s-1}
\label{eq:motzkin_compact}
\end{equation}
where $T(n,k) = \frac{n!}{(n-2k)!\, k!\, (k+1)!}$ are the Motzkin triangle coefficients~\cite{Sloane2023}, and:
\begin{equation}
M_n(t) = \sum_{k=0}^{\lfloor n/2 \rfloor} T(n,k) \cdot t^{n-2k}
\label{eq:motzkin_poly}
\end{equation}

This expression counts all paths that return to the origin after exactly $m_s$ steps, with $r$ representing the probability of a backward step.

\textbf{Physical interpretation.} The Motzkin polynomial encodes three scattering processes: (i) up steps representing positive $z$-projection changes, (ii) down steps representing negative $z$-projection changes (backscattering), and (iii) flat steps representing zero $z$-projection change (forward scattering perpendicular to detection axis).

\subsection{Projection onto one dimension}

Projecting a three-dimensional photon walk onto the surface normal transforms it into a hidden Markov chain: each projected step depends on the previous 3D direction, which is not directly observed. This motivates substituting an effective backscattering coefficient $r_b(g, m_s)$ into the one-dimensional Motzkin expression:
\begin{equation}
P_\mathrm{refl}^{(3D)}(g, m_s) \approx P_\mathrm{refl}^{(1D)}(m_s, r_b(g, m_s))
\end{equation}

The fundamental discovery is that Motzkin polynomial structure persists when embedded in three-dimensional anisotropic scattering, provided we use the correct dimensional reduction parameter transform. This demonstrates that complex multidimensional transport can be preserved in tractable combinatorial representations through proper understanding of boundary constraint physics.

\subsection{Comparison with generalized Kubelka--Munk}

Sandoval and Kim~\cite{SandovalKim2014} pursued an alternative strategy: extending Kubelka--Munk to three spatial dimensions through the double spherical harmonics ($\mathrm{DP}_1$) approximation. Their generalized Kubelka--Munk (gKM) equations form an $8 \times 8$ system governing forward and backward power flow for normally incident collimated beams on finite slabs. For isotropic scattering in optically thick media, gKM achieves quantitative accuracy with errors below 15\% when $z_0 \geq 10$.

However, the gKM approximation exhibits fundamental limitations for anisotropic scattering. The $\mathrm{DP}_1$ expansion employs basis functions containing only first-order azimuthal harmonics, which cannot represent the angular concentration characteristic of forward-peaked phase functions. Sandoval and Kim reported that for Henyey--Greenstein scattering with $g = 0.8$, the gKM solution develops spurious oscillations with errors exceeding 80\% in transmitted power. At higher anisotropy, the approximation yields negative intensities---an unphysical result indicating complete breakdown.

The BTF framework addresses this limitation by inverting the dimensional strategy. Instead of enriching one-dimensional equations with three-dimensional spatial coupling, we project three-dimensional angular physics onto one-dimensional combinatorial structures while compensating for boundary-constrained truncation effects. For the semi-infinite geometry under normal incidence, this achieves mean deviations below 2\% for $g \leq 0.7$, with systematic drift at higher anisotropy that remains physically valid. The key advance is robustness: where gKM fails abruptly at moderate anisotropy, the BTF framework degrades predictably, maintaining utility across the full parameter range relevant to tissue optics and atmospheric physics.

Table~\ref{tab:comparison} summarizes these methodological differences.

\begin{table}[htbp]
\centering
\small
\caption{Comparison of dimensional reduction strategies for normal incidence.}
\label{tab:comparison}
\begin{tabular}{lll}
\hline
\textbf{Property} & \textbf{gKM (Sandoval \& Kim)} & \textbf{BTF (this work)} \\
\hline
Geometry & Finite slab & Semi-infinite \\
Incidence & Normal & Normal \\
Strategy & 1D $\to$ 3D extension & 3D $\to$ 1D projection \\
Angular basis & $\mathrm{DP}_1$ (4 func./hemisphere) & HG sampling + Motzkin \\
Isotropic limit & $<$15\% error for $z_0 \geq 10$ & Exact (Catalan structure) \\
Anisotropic range & $g \lesssim 0.5$ (qualitative) & $g \leq 0.7$ ($<$2\% deviation) \\
High-$g$ behavior & Unphysical (negative values) & Correctable drift \\
Computational form & $8 \times 8$ PDE system & 1D polynomial evaluation \\
\hline
\end{tabular}
\end{table}

\section{Henyey--Greenstein phase function}

For forward-peaked scattering with anisotropy factor $g = \langle\mu\rangle$, successive photon directions remain correlated over multiple events. The Henyey--Greenstein phase function~\cite{HenyeyGreenstein1941}:
\begin{equation}
P_\mathrm{HG}(\mu, g) = \frac{1-g^2}{2(1+g^2-2g\mu)^{3/2}}
\label{eq:hg_phase}
\end{equation}
exhibits the compositional property $P_\mathrm{HG}(g) \otimes P_\mathrm{HG}(g) = P_\mathrm{HG}(g^2)$~\cite{Tretakov2007,PfeifferChapman2008}. Repeated scattering convolves copies of $P_\mathrm{HG}$ and multiplies the anisotropy factors, so that the expectation of the projected path displacement is $\langle\ell\rangle = 1 + g + g^2 + \cdots = 1/(1-g)$.

The cumulative distribution function ($g \neq 0$):
\begin{equation}
F(\mu; g) = \frac{1-g^2}{2g} \left[\frac{1}{\sqrt{1+g^2-2g\mu}} - \frac{1}{1+g}\right]
\label{eq:hg_cdf}
\end{equation}
For $g = 0$: $F(\mu; g) = (1+\mu)/2$.

The inverse CDF used for Monte Carlo sampling ($g \neq 0$):
\begin{equation}
F^{-1}(\xi, g) = \frac{1}{2g}\left[1+g^2 - \left(\frac{1-g^2}{1-g+2g\xi}\right)^2\right]
\label{eq:hg_inverse_cdf}
\end{equation}
For $g = 0$: $F^{-1}(\xi, g) = 2\xi - 1$.

\section{Boundary Truncation Factor framework}

\subsection{Physical formulation}

The mapping from three-dimensional anisotropic transport to one-dimensional combinatorial structures requires understanding how boundary constraints modify scattering behavior. In bulk scattering, the effective asymmetry parameter after $m_s$ steps follows standard multiple scattering theory: $g_\mathrm{eff}^{(\mathrm{bulk})} = g^{m_s}$.

However, first-passage problems~\cite{Redner2001,CondaminBenichou2005} impose geometric return constraints that fundamentally modify this relationship. The requirement that photons return to their entry boundary restricts the accessible angular phase space at each scattering event. This geometric truncation accumulates over multiple scattering events, effectively reducing the asymmetry parameter:
\begin{equation}
g_\mathrm{eff}^{(\mathrm{constrained})} = g^{m_s}\,\mathrm{BTF}(g, m_s)
\end{equation}
where $\mathrm{BTF} \leq 1$ represents the multiplicative reduction in effective anisotropy due to boundary constraints.

BTF emerges from the boundary-constrained angular integration using the phase function (Eq.~\ref{eq:hg_phase}):
\begin{equation}
g_\mathrm{eff}^{(\mathrm{constrained})} = \frac{\int\cdots\int \prod_{i=1}^{m_s} P_\mathrm{HG}(\mu_i, g) \times [\mathrm{constraint}]\, d\mu_1 \cdots d\mu_{m_s}}{\int\cdots\int \prod_{i=1}^{m_s} P_\mathrm{HG}(\mu_i, g)\, d\mu_1 \cdots d\mu_{m_s}}
\end{equation}
where the boundary constraint ensures photons cross below $z = 0$ after exactly $m_s$ scattering events. These integrals become mathematically intractable beyond $m_s = 3$.

\subsection{Monte Carlo determination}

Monte Carlo simulations provide exact first-return probabilities for three-dimensional photon transport~\cite{Jacques2010,Sassaroli1998}. The computational campaign covered $g \in [0.05, 0.95]$ (19 values) and $m_s \in [2, 100]$ (99 values) with $10^8$ trajectories per parameter combination---1,881 parameter combinations repeated across 10 independent runs, yielding approximately 19,000 probability calculations over six months.

BTF is determined as the value needed to make one-dimensional models reproduce Monte Carlo 3D first-return probability:
\begin{equation}
\mathrm{BTF}(g, m_s) = \frac{g_\mathrm{eff}^{(\mathrm{required})}}{g^{m_s}}
\end{equation}
where $g_\mathrm{eff}^{(\mathrm{required})}$ is the effective anisotropy needed to match Monte Carlo results. Analytical solutions exist for $m_s = 2$ and $m_s = 3$, providing exact validation benchmarks.

\subsection{Mathematical representation}

Systematic model selection using Pad\'{e} approximation---starting with high-order polynomials and progressively reducing complexity while monitoring cross-validation---revealed that the optimal functional form exhibits Lorentzian behavior:
\begin{equation}
\mathrm{BTF}(g, m_s) = \frac{\mathrm{norm}(g)}{1 + \left(\frac{m_s - 2}{m_x(g)}\right)^2}
\label{eq:btf_parameterized}
\end{equation}
where:
\begin{align}
\mathrm{norm}(g) &= 1 - \tfrac{1}{2}g(1+g) \quad \text{(normalization factor)} \\
m_x(g) &= \frac{4g}{1-g} \quad \text{(characteristic length scale)}
\end{align}

This parameterized form is algebraically equivalent to the compact expression:
\begin{equation}
\mathrm{BTF}(g, m_s) = \frac{1}{2} \cdot \frac{(1-g)(2+g)}{1 + \left(\frac{g}{1-g}\right)^2(m_s-2)^2}
\label{eq:btf_compact}
\end{equation}

For short paths ($m_s - 2 \ll m_x$), $\mathrm{BTF} \approx \mathrm{norm}(g)$ with minimal boundary effects. For long paths ($m_s - 2 \gg m_x$), $\mathrm{BTF} \to 0$ as boundary truncation dominates. The functional form reproduces Monte Carlo-derived BTF values with mean deviation $<$1\% and cross-validated $R^2 > 0.999$ for $g \leq 0.7$.

\subsection{Theoretical determination of backscattering coefficient}

The backscattering coefficient $r_b(g)$ for the two-step case ($m_s = 2$) is determined through return analysis. The theoretical two-step return probability is:
\begin{equation}
p_{r2}(g) = \int_{\mu=-1}^{\mu=0} \frac{\mu-1}{\mu} \times P_\mathrm{HG}(\mu, g)\, d\mu
\label{eq:pr2}
\end{equation}

The forward/backward threshold $\mu_b(g)$ is determined by matching this benchmark to the one-dimensional first-passage probability at two steps. Using the HG compositional property, the two-step backscattering mass above the threshold satisfies:
\begin{equation}
p_{r2}(g) - \tfrac{1}{2} \times F(-\mu_b; g^2) = 0
\end{equation}

Solving explicitly:
\begin{equation}
\mu_b(g) = -F^{-1}\bigl(2p_{r2}(g), g^2\bigr)
\label{eq:mu_b}
\end{equation}

The final backscattering coefficient:
\begin{equation}
r_b(g) = F(-\mu_b(g); g^2)
\label{eq:rb_final}
\end{equation}

In practice, $p_{r2}(g)$ is evaluated analytically and $\mu_b(g)$ is obtained by a one-dimensional root finder, establishing $r_b$ purely from theory without Monte Carlo input.

\subsection{Computational algorithm for \texorpdfstring{3D$\to$1D}{3D to 1D} mapping}

The complete parameter transformation proceeds in four steps:

\textbf{Step 1 (Angular threshold):} Compute $\mu_b(g)$ from Eq.~\ref{eq:mu_b}.

\textbf{Step 2 (Dimensional reduction exponent):} Compute the effective exponent:
\begin{equation}
\gamma(g, m_s) = 2 + \mathrm{BTF}(g, m_s)(m_s - 2)
\label{eq:gamma}
\end{equation}

\textbf{Step 3 (Effective backscattering):} Substitute into the HG CDF:
\begin{equation}
r_b(g, m_s) = F\bigl(-\mu_b(g); g^{\gamma(g, m_s)}\bigr)
\label{eq:rb_substitution}
\end{equation}

\textbf{Step 4 (First-return probability):} Evaluate the Motzkin polynomial expression (Eq.~\ref{eq:motzkin_compact}) with the substituted $r_b$:
\begin{equation}
P_\mathrm{3D}^{(\mathrm{refl})}(g, m_s) = P_\mathrm{1D}^{(\mathrm{refl})}\bigl(m_s, r_b(g, m_s)\bigr)
\end{equation}

\textbf{Limiting case (ballistic propagation):} In the absence of early returns, angular correlations are perfectly preserved with $g_\mathrm{eff} = g^{m_s}$. The BTF reduces to unity, hence $\gamma(g, m_s) = m_s$, and the framework correctly reduces to standard multiple-scattering theory.

\section{Monte Carlo validation}

We define first return as the event where a photon's cumulative $z$-coordinate first becomes negative within $m_s$ scattering steps. Each trajectory consists of $m_s$ steps with exponentially distributed lengths and angular deflections sampled from the Henyey--Greenstein phase function.

\textbf{Monte Carlo procedure:}
\begin{enumerate}
\item Initialize photon at origin ($z = 0^+$)
\item For each step $i = 1$ to $m_s$:
\begin{itemize}
\item Sample step length from exponential distribution (unit mean free path)
\item Sample scattering angle from HG phase function using $F^{-1}$ (Eq.~\ref{eq:hg_inverse_cdf})
\item Update position and check if $z < 0$
\end{itemize}
\item Record first occurrence of $z < 0$ as successful return
\item Repeat for $10^8$ trajectories per parameter combination
\end{enumerate}

\begin{figure}[H]
	\centering
	\includegraphics[width=\textwidth]{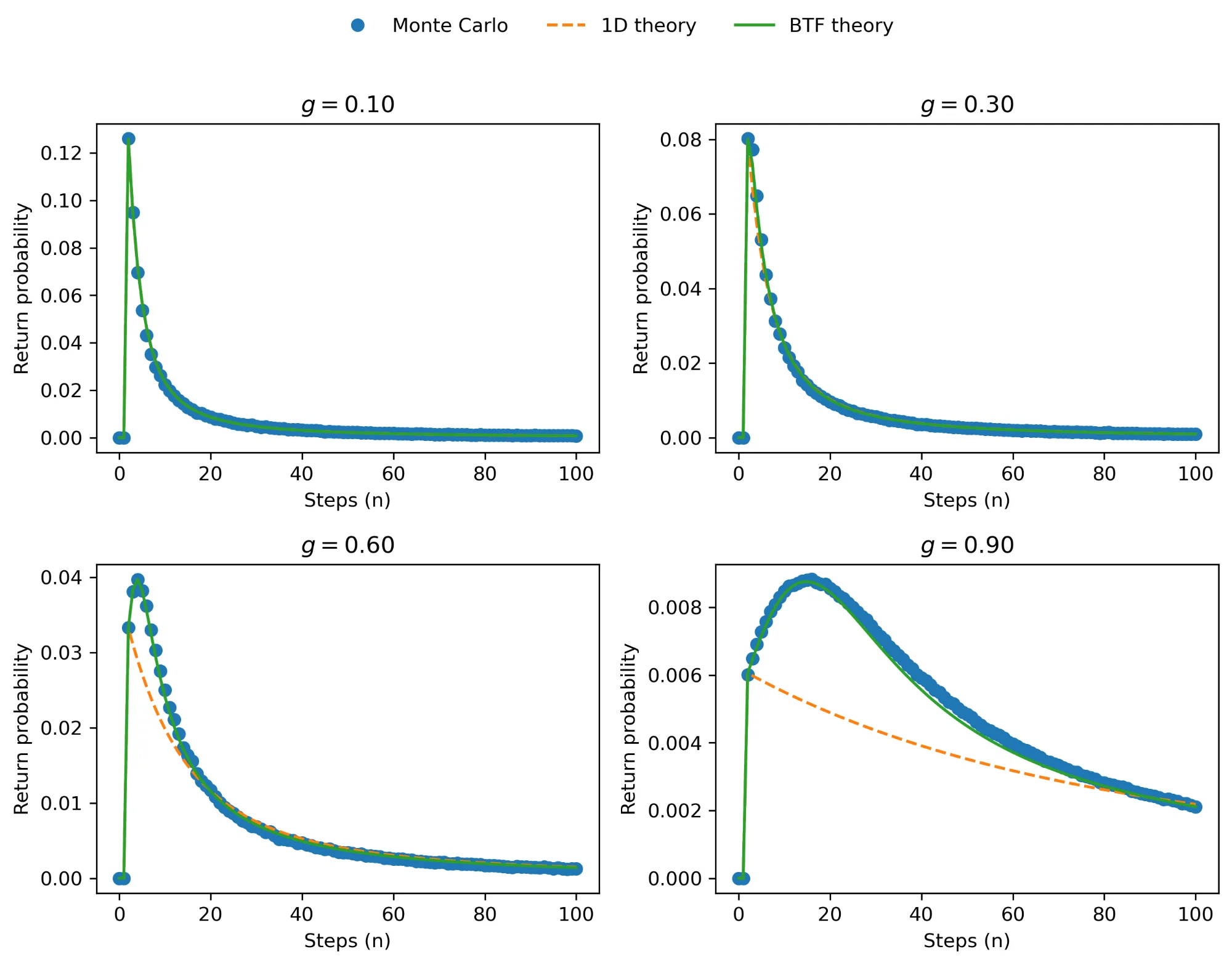}
	\caption{Comparison of Monte Carlo data with old and new theoretical fits.}
	\label{fig:return-prob-fits}
\end{figure}

The dimensional reduction framework demonstrates internal consistency by reproducing analytical solutions exactly for $m_s = 2$ and $m_s = 3$. It exhibits physically reasonable scaling behavior across parameter space and maintains statistical equivalence between 3D and 1D representations. Multiple independent simulation campaigns with different code implementations confirm reproducibility.

\section{Conclusions}

This paper establishes a hybrid analytical-numerical benchmark for radiative transport in semi-infinite media with Henyey--Greenstein scattering. Four principal contributions emerge:

\textbf{1. Dimensional reduction with preserved combinatorics.} The mapping from 3-D anisotropic transport to 1-D Motzkin polynomials (Eq.~\ref{eq:motzkin_poly})~\cite{OsteVanderJeugt2015} is exact for first-passage and first-return statistics. The axial projection constitutes a hidden Markov process enabling tractable computation while preserving essential physics.

\textbf{2. Boundary Truncation Factor (BTF).} The BTF (Eq.~\ref{eq:btf_compact}) provides a systematic approach to analytically intractable integrals at higher scattering orders. The striking result that BTF values follow a Cauchy (Lorentzian) distribution remains unexplained and warrants theoretical investigation.

\textbf{3. Robust statistical equivalence.} Monte Carlo validation across $g \in [0.05, 0.95]$ and $m_s \in [2, 100]$ confirms mean deviations below 2\% for $g \leq 0.7$, with predictable drift at higher anisotropy. Backscattering coefficients are derived from phase-function integrals without empirical fitting.

\textbf{4. Universal scaling laws.} Discovery of scaling relationships governing parameter transformation across the transition from early-time enhancement to asymptotic diffusive decay.

\subsection{Limitations}

The current framework is restricted to Henyey--Greenstein phase functions, planar boundary geometry, normal incidence illumination, and empirical rather than first-principles BTF derivation.

\subsection{Future directions}

Extensions merit investigation: theoretical derivation of BTF from first principles; generalized phase functions (Gegenbauer, two-term Henyey--Greenstein~\cite{PfeifferChapman2008}, Rayleigh); oblique incidence and broad beam illumination; complex geometries (spherical, cylindrical boundaries); experimental validation; and polarization effects.

The practical payoff is computational efficiency: the framework replaces expensive 3-D Monte Carlo simulations with 1-D polynomial evaluation, enabling iterative algorithms in tissue optics~\cite{JacquesTissue2013}, remote sensing, and material appearance modeling.

\section*{Acknowledgements}

The author (1) thanks Dr.\ Florence Zeller for illuminating discussions on Chebyshev polynomials and Motzkin combinatorial structures. The author (1) is also grateful to Professor Arnold Kim for encouragement and venue guidance, and to Professor Michel Talagrand for his honest perspective on the difficulty of proving the observed Cauchy distribution in the BTF. The authors used AI tools for editorial assistance in preparing this manuscript.


\begin{thebibliography}{99}

\bibitem{HenyeyGreenstein1941}
L.~G.~Henyey and J.~L.~Greenstein,
``Diffuse radiation in the galaxy,''
\textit{Astrophys.\ J.}\ \textbf{93}, 70--83 (1941).

\bibitem{Chandrasekhar1960}
S.~Chandrasekhar,
\textit{Radiative Transfer}
(Dover Publications, 1960).

\bibitem{Chandrasekhar1943}
S.~Chandrasekhar,
``Stochastic problems in physics and astronomy,''
\textit{Rev.\ Mod.\ Phys.}\ \textbf{15}, 1--89 (1943).

\bibitem{KubelkaMunk1931}
P.~Kubelka and F.~Munk,
``An article on optics of paint layers,''
\textit{Z.\ Tech.\ Phys.}\ \textbf{12}, 593--601 (1931).

\bibitem{SandovalKim2014}
C.~Sandoval and A.~D.~Kim,
``Deriving {Kubelka--Munk} theory from radiative transport,''
\textit{J.\ Opt.\ Soc.\ Am.\ A}\ \textbf{31}, 628--636 (2014).

\bibitem{SandovalKim2017}
C.~Sandoval and A.~D.~Kim,
``Generalized {Kubelka--Munk} approximation for multiple scattering of polarized light,''
\textit{J.\ Opt.\ Soc.\ Am.\ A}\ \textbf{34}, 153--162 (2017).

\bibitem{OsteVanderJeugt2015}
R.~Oste and J.~Van~der~Jeugt,
``Motzkin paths, {Motzkin} polynomials and recurrence relations,''
\textit{Electron.\ J.\ Combin.}\ \textbf{22}(2), P2.8 (2015).

\bibitem{ZellerCordery2020}
C.~Zeller and R.~Cordery,
``Light scattering as a {Poisson} process and first-passage probability,''
\textit{J.\ Stat.\ Mech.}\ \textbf{2020}, 063404 (2020).

\bibitem{Jacques2010}
S.~L.~Jacques,
``{Monte Carlo} modeling of light transport in tissue (steady state and time of flight),''
in \textit{Optical-Thermal Response of Laser-Irradiated Tissue}, 2nd ed.
(Springer, 2010), pp.\ 109--144.

\bibitem{Sassaroli1998}
A.~Sassaroli, C.~Blumetti, F.~Martelli, L.~Alianelli, D.~Contini, A.~Ismaelli, and G.~Zaccanti,
``{Monte Carlo} procedure for investigating light propagation and imaging of highly scattering media,''
\textit{Appl.\ Opt.}\ \textbf{37}, 7392--7400 (1998).

\bibitem{JacquesTissue2013}
S.~L.~Jacques,
``Optical properties of biological tissues: a review,''
\textit{Phys.\ Med.\ Biol.}\ \textbf{58}, R37--R61 (2013).

\bibitem{Modric2009}
D.~Modri\'{c}, S.~Bolan\v{c}a, and R.~Beuc,
``{Monte Carlo} modeling of light scattering in paper,''
\textit{J.\ Imaging Sci.\ Technol.}\ \textbf{53}, 020201 (2009).

\bibitem{Ballestra2016}
L.~V.~Ballestra, G.~Pacelli, and D.~Radi,
``A very efficient approach to compute the first-passage probability density function in a time-changed {Brownian} model: Applications in finance,''
\textit{Physica A}\ \textbf{463}, 330--344 (2016).

\bibitem{YangKruse2004}
L.~Yang and B.~Kruse,
``Revised {Kubelka--Munk} theory.\ {I}.\ {Theory} and application,''
\textit{J.\ Opt.\ Soc.\ Am.\ A}\ \textbf{21}, 1933--1941 (2004).

\bibitem{Haney2007}
M.~M.~Haney and K.~van~Wijk,
``Modified {Kubelka--Munk} equations for localized waves inside a layered medium,''
preprint arXiv:physics/0701193 (2007).

\bibitem{Myrick2011}
M.~L.~Myrick, M.~N.~Simcock, M.~Baranowski, H.~Brooke, S.~L.~Morgan, and J.~N.~McCutcheon,
``The {Kubelka--Munk} diffuse reflectance formula revisited,''
\textit{Appl.\ Spectrosc.\ Rev.}\ \textbf{46}, 140--165 (2011).

\bibitem{PhilipsInvernizzi2001}
B.~Philips-Invernizzi, D.~Dupont, and C.~Caze,
``Bibliographical review for reflectance of diffusing media,''
\textit{Opt.\ Eng.}\ \textbf{40}, 1082--1092 (2001).

\bibitem{CondaminBenichou2005}
S.~Condamin, O.~B\'{e}nichou, and M.~Moreau,
``First-passage times for random walks in bounded domains,''
\textit{Phys.\ Rev.\ Lett.}\ \textbf{95}, 260601 (2005).

\bibitem{BlancoFournier2006}
S.~Blanco and R.~Fournier,
``Short-path statistics and the diffusion approximation,''
\textit{Phys.\ Rev.\ Lett.}\ \textbf{97}, 230604 (2006).

\bibitem{Libois2022}
Q.~Libois and A.~B.~Davis,
``Photon path distributions in optically thin slabs,''
\textit{Opt.\ Express}\ \textbf{30}, 40968 (2022).

\bibitem{BorovkovBorovkov2008}
A.~A.~Borovkov and K.~A.~Borovkov,
\textit{Asymptotic Analysis of Random Walks: Heavy-Tailed Distributions}
(Cambridge University Press, 2008).

\bibitem{McDonald1999}
D.~R.~McDonald,
``Asymptotics of first passage times for random walk in an orthant,''
\textit{Ann.\ Appl.\ Probab.}\ \textbf{9}(1), 110--145 (1999).

\bibitem{Redner2001}
S.~Redner,
\textit{A Guide to First-Passage Processes}
(Cambridge University Press, 2001).

\bibitem{RudnickGaspari2004}
J.~Rudnick and G.~Gaspari,
\textit{Elements of the Random Walk}
(Cambridge University Press, 2004).

\bibitem{DrakeGantner2011}
D.~Drake and R.~Gantner,
``Generating functions for plateaus in {Motzkin} paths,''
preprint arXiv:1109.3272 (2011).

\bibitem{SimonTrachsler2003}
K.~Simon and B.~Trachsler,
``A random walk approach for light scattering in material,''
\textit{Discrete Math.\ Theor.\ Comput.\ Sci.}, 289--300 (2003).

\bibitem{Sloane2023}
N.~J.~A.~Sloane,
``A handbook of integer sequences fifty years later,''
preprint arXiv:2301.03149 (2023).

\bibitem{Tretakov2007}
V.~M.~Petnikova, V.~V.~Shuvalov, and E.~V.~Tret'akov,
``Multiple-scattering {Henyey--Greenstein} phase function and fast path-integration,''
in \textit{Int.\ Conf.\ Coherent and Nonlinear Optics} (SPIE, 2007), p.\ 672726.

\bibitem{PfeifferChapman2008}
N.~Pfeiffer and G.~H.~Chapman,
``Successive order, multiple scattering of two-term {Henyey--Greenstein} phase functions,''
\textit{Opt.\ Express}\ \textbf{16}, 13637--13645 (2008).

\end{thebibliography}
\end{document}